\documentclass[10pt, two column, twoside]{IEEEtran}
\usepackage{color}
\usepackage[colorlinks, linkcolor=color1, anchorcolor=blue, citecolor=color1]{hyperref}

\usepackage{amssymb}
\usepackage{amsmath}
\usepackage[lined,boxed,commentsnumbered, ruled]{algorithm2e}
\usepackage{mathrsfs}
\usepackage{algorithmic}
\usepackage{bm}
\usepackage{hhline}

\usepackage{pgfplots}
\usepackage{tikz}
\usetikzlibrary{arrows}
\usepackage{subfigure}
\usepackage{graphicx,booktabs,multirow}

\definecolor{colorhkust}{RGB}{20,43,140}
\definecolor{colortsinghua}{RGB}{116,52,129}
\definecolor{color1}{RGB}{128,0,0}

\date{}

\begin{document}

        \title{The Roadmap to 6G -- AI Empowered \\
        Wireless Networks}
\author{Khaled B. Letaief, Wei Chen, Yuanming~Shi, Jun Zhang, and Ying-Jun
Angela Zhang

        \thanks{Khaled B. Letaief is with Hong Kong University of Science and Technology; Wei Chen is with Tsinghua University; Yuanming Shi is with ShanghaiTech University; Jun Zhang is with The Hong Kong Polytechnic University; Ying-Jun Angela Zhang is with The Chinese University of Hong Kong. }
\thanks{This work has been submitted to the IEEE for possible publication. Copyright may be transferred without notice, after which this version may no longer be accessible.}
                }

\maketitle


\maketitle

\begin{abstract}
The recent upsurge of diversified mobile applications, especially those supported by Artificial Intelligence (AI), is spurring heated discussions on the future evolution of wireless communications. While 5G is being  deployed around the world, efforts from industry and academia have started to look beyond 5G and conceptualize 6G. We envision 6G to undergo an unprecedented transformation that will make it substantially different from the previous generations of wireless cellular systems. In particular, 6G will go beyond mobile Internet and will be required to  support ubiquitous AI services from the core to the end devices of the network. Meanwhile, AI will  play a critical role in designing and optimizing 6G architectures, protocols, and operations. In this article, we discuss potential technologies for 6G to enable mobile AI applications, as well as AI-enabled methodologies for 6G network design and optimization. Key trends in the evolution to 6G will also be discussed.


\end{abstract}


\section{Introduction}
\begin{figure*}[t]
\center
\includegraphics[width=2\columnwidth]{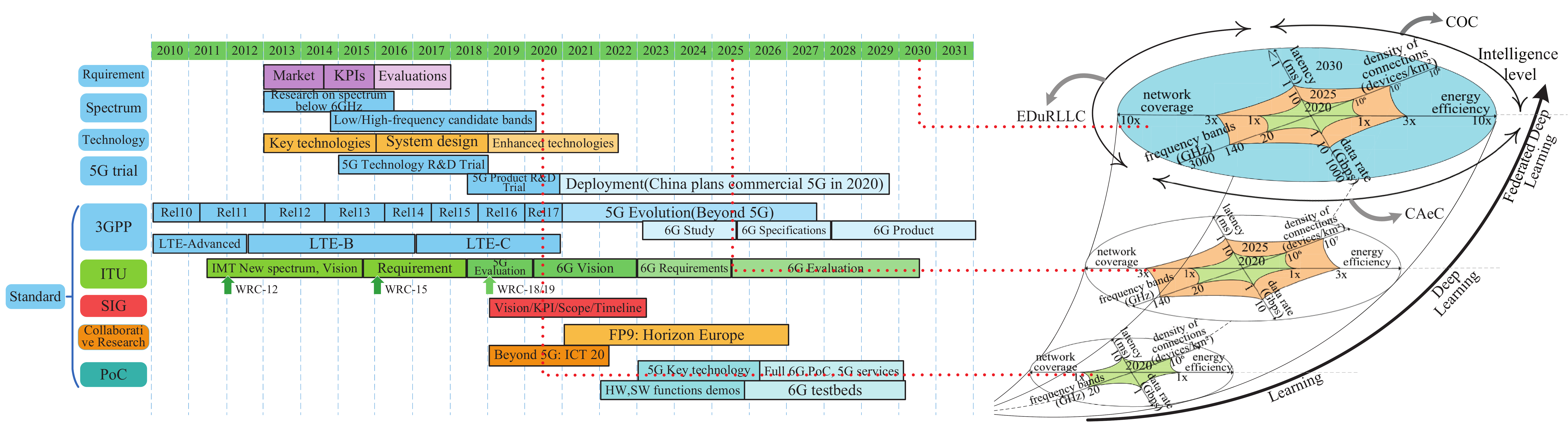}
\caption{The roadmap of 6G.}
\label{6Groad}
\end{figure*}

The wireless communications industry is one of the few industry sectors that
have kept a fast growing trend with creative features for
a number of decades. The current 4G LTE networks have led to the thriving of mobile Internet,
enabling various innovative applications, such as mobile shopping and payment,
smart home/city, mobile gaming, etc. The great success of mobile Internet
has in turn been a driving force behind the evolution of wireless technologies. The upcoming 5G
network will support a wide range of services, including eMBB (enhanced mobile
broadband), uRLLC (ultra-reliable and low-latency communications), and mMTC
(massive machine-type communications) \cite{Fettweis_lowlatency18,Jeff_JSAC5G}. According to a Cisco forecast,
 major operators will embark on a significant investment in 5G networks during the next one or two years.

While 5G is still at an initial stage, to maintain the sustainability and competitiveness of wireless communication systems,
it is time for both the industry and academia to think about what 6G will
be. There are already initiatives describing the roadmap towards 6G \cite{Berndt_6G, strinati20196g,
Debbah_arXiv196G} along with the emerging trends and requirements, as well as various enabling
techniques and architectures, e.g., Terahertz band communications \cite{Rappaport_Globecom18}.

In contrast to previous generations, 6G will be transformative and will revolutionize the wireless evolution from ``connected things" to ``connected intelligence" with more stringent requirements specified as follows.
\begin{itemize}
\item Very high data rates, up to 1 Tbps;
\item Very high energy efficiency, with the ability to support battery-free IoT devices;
\item Trusted global connectivity;
\item Massive low-latency control (less than 1 msec end-to-end latency);
\item Very broad frequency bands (e.g., 73GHz-140GHz and 1THz-3THz);
\item Ubiquitous always-on broadband global network coverage by integrating terrestrial wireless with satellite systems;
\item Connected intelligence with machine learning capability and AI networking hierarchy.
\end{itemize}

6G will also require the support of three new service types beyond the eMBB, uRLLC,
and mMTC services supported by 5G, as described below.

\textbf{Computation Oriented Communications (COC):} New smart
devices  call for distributed and in-network computation to enable the key
functionalities of AI-empowered 6G, such as federated learning and edge intelligence. Instead of targeting
classical quality of service (QoS) provisioning, CoC will flexibly choose
an operating point in the rate-latency-reliability space depending on the
availability of various communications resources to achieve a certain computational
accuracy.

\textbf{Contextually Agile eMBB Communications (CAeC):} The provision of 6G
eMBB services is expected to be more agile and adaptive to the network context, including communication network context such as link congestion
and network topology; physical environment context such as surrounding location
and mobility; and social network context such as social neighborhood and
sentiments.

\textbf{Event Defined uRLLC (EDuRLLC):} In contrast to the 5G uRLLC application scenario
(e.g., virtual reality and industrial automation) where redundant resources
are in place to offset many uncertainties, 6G will need to support uRLLC in extreme
or emergency events with spatially and temporally changing device densities,
traffic patterns, and spectrum and infrastructure availability.

Inspired by these trends, in this article, we attempt to conceptualize 6G as an intelligent information
system that is both driven by and a driver of the modern AI technologies. A roadmap for 6G is depicted in Fig. {\ref{6Groad}}, which is plotted based on the strategic plans of various standard bodies and is also projected based on the 5G status. Key performance indicators (KPIs) and service types are also illustrated. Meanwhile, a potential network architecture for 6G is shown in Fig. {\ref{6G}}. We envision that AI will greatly enhance the situational awareness of the network operators, and enable close-loop optimization to support the new service types as mentioned above. As such, 6G will unleash the full potential of mobile communications, computing, and control in a host of exciting applications, including
smart cities, connected infrastructure, wearable computers, autonomous driving,
UAVs \cite{Gesbert_IoT19}, seamless virtual and augmented reality, Internet of Things, space-air-ground integrated networks \cite{Kato_WC19AI}, and a lot more.

This article is a humble attempt to provide a forward-looking research roadmap for 6G. The rest of the article is organized as follows. In Section \ref{AInet}, a vision on 6G architecture will be presented. In Section \ref{AIcom}, we will show how 6G leverages the advent of AI to enable its key features. Various AI applications of 6G will be given in Section \ref{AIapp}. To meet the expected stringent requirements of such applications, hardware-aware communications will be embraced in 6G, as discussed in Section \ref{hardwarecom}. Finally, Section \ref{conclud} concludes the article.

\section{The Architecture of 6G Networks}
\label{AInet}
\begin{figure*}[t]
\center
\includegraphics[scale = 0.33]{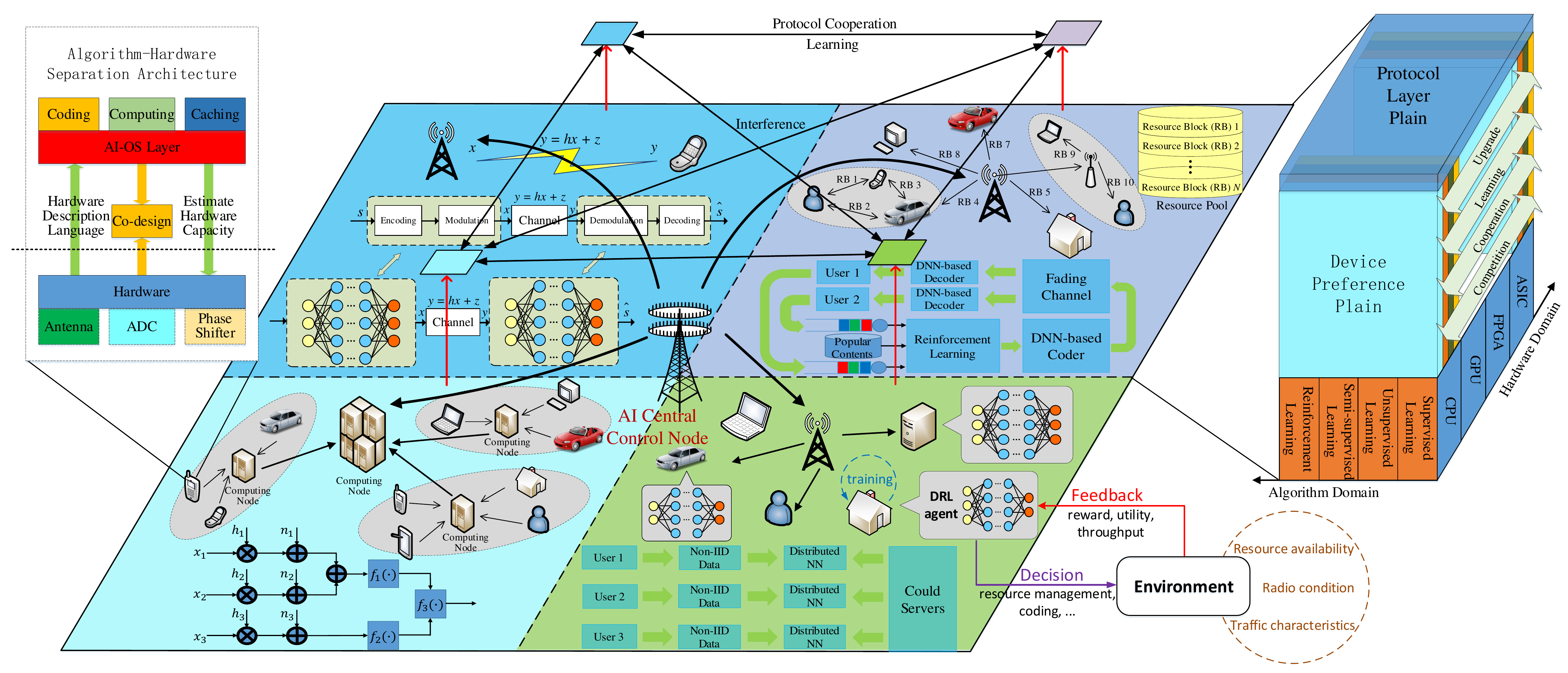}
\caption{The architecture of 6G.}
\label{6G}
\end{figure*}


In this section, we introduce a potential architecture for 6G as shown in Fig. {\ref{6G}}, in which network intelligentization, subnetwork evolution, and intelligent radio are embraced.
\subsection{From Network Softwarization to Network Intelligentization}
We envision that 6G will take network softwarization to a new level, namely, towards network intelligentization. In 5G, the ``non-radio'' aspect has become more and more important, and has been the key driver behind  the recent efforts on  ``softwarization''. More specifically, two key 5G technologies are Software-Defined Networking (SDN) and Network Functions Virtualization (NFV), which have moved modern communications networks towards software-based virtual networks.  They also enable  network slicing, which can provide a powerful virtualization capability to allow multiple virtual networks to be created atop a shared physical infrastructure.

Nevertheless, as the network is becoming more complex and more heterogeneous, softwarization is not going to be sufficient for beyond 5G networks.  In particular, to support AI-based applications, the network entities have to support diverse capabilities, including communications, content caching, computing, and even wireless power transfer. Furthermore, 6G will embrace new radio access interfaces such as THz communications and intelligent surfaces. It will also need to support  more advanced Internet of Things (IoT) functionalities including sensing, data collection, analytics, and storage. All of the aforementioned challenges call for an architecture that is flexible, adaptive, and more importantly, intelligent. Existing technologies, such as SDN, NFV, and network slicing will need  to be further improved to meet these challenges. By enabling fast learning and adaptation, AI-based methods will render network slicing a lot more versatile in 6G systems.

The design of the 6G architecture shall follow an ``AI native'' approach where intelligentization will allow the network to be smart, agile, and able to learn and adapt itself according to the changing network dynamics. It will evolve into a ``network of subnetworks,'' allowing more efficient and flexible upgrades, and a new framework based on intelligent radio and algorithm-hardware separation  to cope with the heterogeneous and upgradable hardware capabilities. Both of these two features will exploit AI techniques, as further illustrated in the following subsections.

\subsection{A Network of Subnetworks -- Local vs Global Evolution}
Given its expected ultra-high heterogeneity, one key feature of 6G will be its capability to exploit a flexible subnetwork-wide evolution to effectively adapt to the local environments and user demands, thereby resulting in a ``network of subnetworks''. Particularly, local subnetworks in 6G may evolve individually to upgrade themselves. The local evolution may happen in a few neighboring cells or even in a single cell in order to flexibly apply cutting-edge developments on new waveforms, coding, and multi-access protocols in subnetworks without extensive time-consuming tests. Since there is no need to rebuild the whole system, the evolution cost can  be substantially reduced. To achieve this goal, we need to address the following three challenges:

\begin{enumerate}
\item Each subnetwork should collect and analyze its local data, which may include wireless environments, user requests, mobility patterns, etc. and then exploit AI methods to upgrade itself locally and dynamically.

\item When the local PHY or MAC protocols are changed, the inter-subnetwork interaction is expected to maintain new inter-subnetwork coordination. One possible solution is to adopt game and learning approaches in 6G, which can assure the convergence of the subnetworks upgrades.

\item The local evolution of 6G requires a relatively stable control plane to support the evolution in the ``network of subnetworks'' level. One possible solution relies on the ``learning from scratch'' method developed in Alpha Zero \cite{silver2017mastering}. The control plane of 6G should evaluate each upgrade of subnetworks, and then implement a network-level learning process to identify the best strategy for each subnetwork, accounting for its local environments and user behaviors.

\end{enumerate}

In summary, the local evolution of subnetworks substantially speeds up the deployment of novel physical and MAC layer protocols, and can better adapt to the spatially and temporally varying radio environments and user demands. With the subnetwork-wide upgrades, we envision a smooth evolution from 5G to 6G and beyond.

\subsection{Towards Intelligent Radio (IR)}
\begin{table*}[!t]\footnotesize
        \renewcommand{\arraystretch}{1.2}
        \caption{The comparison of Software Defined Radio (SDR), Cognitive Radio (CR), and Intelligent Radio (IR).}
        \label{table1}
        \centering
        \begin{tabular}{|l||c|c|c|}
                \hline
                & SDR & CR & IR \\
                \hhline{|=||=|=|=|}
                Frequency Band & Fixed & Adapt to environment & Adapt to
                environment and hardware \\
                \hline
                Spectrum Sharing & Fixed & Opportunistic & AI-enabled \\
                \hline
                Hardware Capability & Pre-claimed & Pre-claimed & Online estimated \\
                \hline
                Hardware Upgradability & No & No & Yes \\
                \hline
                PHY Tx/Rx Module & { Modulation/coding/detection/estimation}
                & {Modulation/coding/detection/estimation} & Deep neural networks \\
                \hline
                Multiple Access & Predetermined & Sensing Based & Distributed ML based \\
                \hline
                Protocols over Layer 3 & Fixed & Fixed & Self-upgradable\\
                \hline
                Main Steam Apps & Voice, Data & Multimedia, Data & AI, In-network Computation\\
                \hline
        \end{tabular}
        \vspace*{-15pt}
\end{table*}

The emerging hardware revolutions, e.g., in the RF and circuit systems, will drive 6G to track and fully exploit the fast upgrade of the device-level and base-station level hardware. We envision that an algorithm-hardware separation architecture will become essential in 6G. Particularly, a transceiver algorithm will be able to automatically estimate the capability of the transceiver hardware over which the protocol runs, then configures itself based on the hardware capability.

This is in contrast to the systems from 1G to 5G where the  devices and transceiver algorithms are jointly designed. Conventionally, the hardware capabilities, e.g., the number of antennas, RF chains, and phase shifters, the resolution and sampling rates of ADCs, as well as, the computation abilities of decoders, etc., have remained quasi-static in the previous cellular generations.  However, the recent state-of-the-art circuits and antennas advances are speeding up and significantly improving the hardware capabilities, which make it possible for the 6G  BS and handset to be diversified and upgradable within 6G. In other words, 6G will not be operating under the conventional joint design, which fails in allowing agile adaptation to a  diversified and upgradable hardware.

To overcome the shortcoming of joint hardware-algorithm design and reap the benefit of the algorithm-hardware separation architecture, we present an operating system (OS) between the device hardware and the transceiver algorithms, where we can regard a transceiver algorithm as a software running over the OS. The OS is capable of not only estimating the capabilities of local RF chains, phase shifters, ADCs, and antennas, etc., but also measuring their analog parameters automatically. Based on the hardware information and AI methods, the OS will then be capable of configuring its own transceiver algorithms via an interface language. We shall refer to this framework as  intelligent radio (IR). In contrast to the learning based intelligent PHY layer surveyed in subsection \ref{iwc}, IR is a much broader concept relying on the algorithm-hardware separation architecture. In Table \ref{table1}, we compare key features of IR, software-defined radio (SDR), and cognitive radio. Owing to Mitola's milestone works \cite{Mitola_IPC19}, IR can be regarded as a further extension, in which the cutting edge AI  techniques are deeply involved. The conventional modulation/coding modules are replaced by deep neural networks, which can in an intelligent way adapt to the environment and hardware. IR also takes into account the protocols over layer 3, which are self-upgradable to support various AI applications.

By exploiting IR, 6G is  expected to evaluate the contributions of various hardware components and identify their bottlenecks. In return, the bottleneck analysis helps the device manufactures in optimizing the budget allocation of the hardware costs. As a result, the application of IR will help 6G  enjoy a much reduced implementation time and a significant reduction in the cost  of new algorithms and hardware in both the PHY and MAC layers, thereby speeding up its own evolution.

\section{AI-Enabled Technologies for 6G}
\label{AIcom}
The unprecedented transformation of wireless networks will make 6G substantially different from the previous generations, as it will be characterized by a high degree of heterogeneity in multiple aspects, such as network infrastructures, radio access technologies, RF devices, computing and storage resources, application types, etc.  In addition, the wide range of new applications will mandate an intelligent use of communications, computing, control, and storage resources from the network edge to the core, and across multiple radio technologies and network platforms. Last but not least, the volume and variety of data generated in wireless networks are growing significantly. This opens up great opportunities for data-driven network planning and operation to achieve real-time additivity to dynamic network environments.

In this section, we advocate AI as an indispensable tool to facilitate intelligent learning, reasoning, and decision making in 6G wireless networks.

\subsection{Big Data Analytics for 6G}
\label{bigdata}
The first natural application of AI is big data analytics. There are four types of analytics that can be applied to 6G systems, namely descriptive analytics, diagnostic analytics, predictive analytics, and prescriptive analytics. Descriptive analytics mine historical data to get insights on network performance, traffic profile, channel conditions, user perspectives, and etc.. It greatly enhances the situational awareness of network operators and service providers. Diagnostic analytics enable autonomous detection of network faults and service impairments, identify the root causes of network anomalies, and ultimately improve the reliability and security of 6G wireless systems. Predictive analytics use data to predict future events such as traffic patterns, user locations, user behavior and preference, content popularity, and resource availability. Prescriptive analytics take advantage of the predictions to suggest decision options for resource allocation, network slicing and virtualization, cache placement, edge computing, autonomous driving, etc. For example, by predicting, anticipating, and inferring future user demands through big data analytics, the notion of proactive caching has recently emerged to significantly relieve peak traffic loads from the wireless core network.

\subsection{AI-enabled Closed-loop Optimization}
\label{aiopt}
Traditional methodologies for wireless network optimization may not be applicable in 6G systems due to the following reasons.  First, 6G wireless systems will be extremely dynamic and complex due to the scale, density, and heterogeneity of the network. Modeling such systems is very hard, if not impossible. As such, traditional optimization approaches that rely heavily on mathematically convenient models will no longer be adequate.  Hence, the second major application of AI in 6G wireless systems is automated and closed-loop optimization. Problems in wireless networks are traditionally solved by applying sets of rules derived from system analysis with prior domain knowledge and experience. For example, in traditional network optimization problems, the objective functions are assumed to be  available in nice algebraic forms, allowing an optimizer to evaluate a solution by simple calculation. However, in the complex 6G network environment, the mapping between a decision and its effect on the physical system is cost prohibitive to define and may not be analytically available. Recent advances in AI technologies, such as reinforcement learning and deep reinforcement learning (DRL), can establish a feedback loop between the decision maker and the physical system, so that the decision maker can iteratively refine its action based on the system's feedback to reach optimality eventually. For example, \cite{Liang_DRLarXiv18} recently applied DRL to address several emerging issues in communication and networking, including adaptive modulation, wireless caching, data offloading, and so on, as shown in Fig. {\ref{6G}}.

\subsection{Intelligent Wireless Communication}
\label{iwc}
AI technologies will play a critical role in end-to-end optimization of the full chain of the physical layer signal processing, from the transmitter to the receiver. The end-to-end communication system suffers from a wide variety of impairments, including hardware impairments such as amplifier distortion, quadrature imbalance, local oscillator and clock harmonic leakage, and the channel impairments such as fading and interference. Meanwhile, the number of factors and parameters to be controlled will continue to increase. With this level of complexity, end-to-end optimization has never been practical in today's wireless systems. Instead, existing approaches divide the full chain into multiple independent blocks, each with a simplified model that does not accurately or holistically capture the features of real-world systems.

AI technologies open up the possibilities to learn the best way to communicate over combinations of hardware and channel effects. We envision an ``intelligent PHY layer'' paradigm in 6G, where the end-to-end system is capable of self learning and self optimization by combining advanced sensing and data collection, AI technologies, and domain-specific signal processing approaches. 

\section{6G for AI Applications}
\label{AIapp}
With the ubiquitousness of smart mobile gadgets and the revival of artificial
intelligence, various AI-empowered mobile applications are emerging. In this section, we present how 6G will handle mobile AI applications.

\subsection{Trends and Challenges}

AI has achieved remarkable successes in many application domains, e.g.,
computer vision, natural language processing, and autonomous driving. AI tasks are  computationally intensive and  mostly trained, developed, and deployed at data centers with custom-designed servers.
Given the fast growth of smart mobile gadgets and Internet of Things devices, it is expected that a large number of intelligent applications will be deployed at the edge of wireless networks in the near future. As such, the 6G wireless network will be designed to leverage advanced wireless communications and mobile computing technologies to support AI-enabled applications at various edge mobile devices with limited communication, computation, hardware and energy resources. Notably, the capacity and latency of wireless links are the key bottlenecks of mobile AI applications due to three reasons. First, to protect privacy, some AI applications require
 data to be kept at the mobile devices instead of being uploaded to the cloud
during
the model training process. This has stimulated the recent research interest
for on-device distributed training, i.e., federated learning
 \cite{mcmahan2017communication}, where frequent communications among the
computing devices are needed for model updates. Secondly, to overcome the resource limitation of edge devices,
on-device distributed computing provides new opportunities by pooling the
computation and storage resources of multiple mobile devices.
In this case, data shuffling is a key component for exchanging the computed intermediate
values among mobile devices to enable on-device distributed inference \cite{Yuanming_distcomp18icassp}.
 Last but not least, the heterogeneous mixture
of the cloud, edge and end computing devices provides a dispersed computing
environment for both training and inference of deep neural networks.

To  enable ubiquitous
and diversified mobile {AI services}, 6G is expected to provide flexible platforms for developing advanced
communication and computation technologies. Moreover, it will provide a holistic way to optimize across the communication, computation, and storage resources to span the functionalities of  modern AI across the end-devices, network edges, and cloud data centers.

\subsection{Communication for Distributed Machine Learning}
\begin{figure}[t]
\center
\includegraphics[width=0.85\columnwidth]{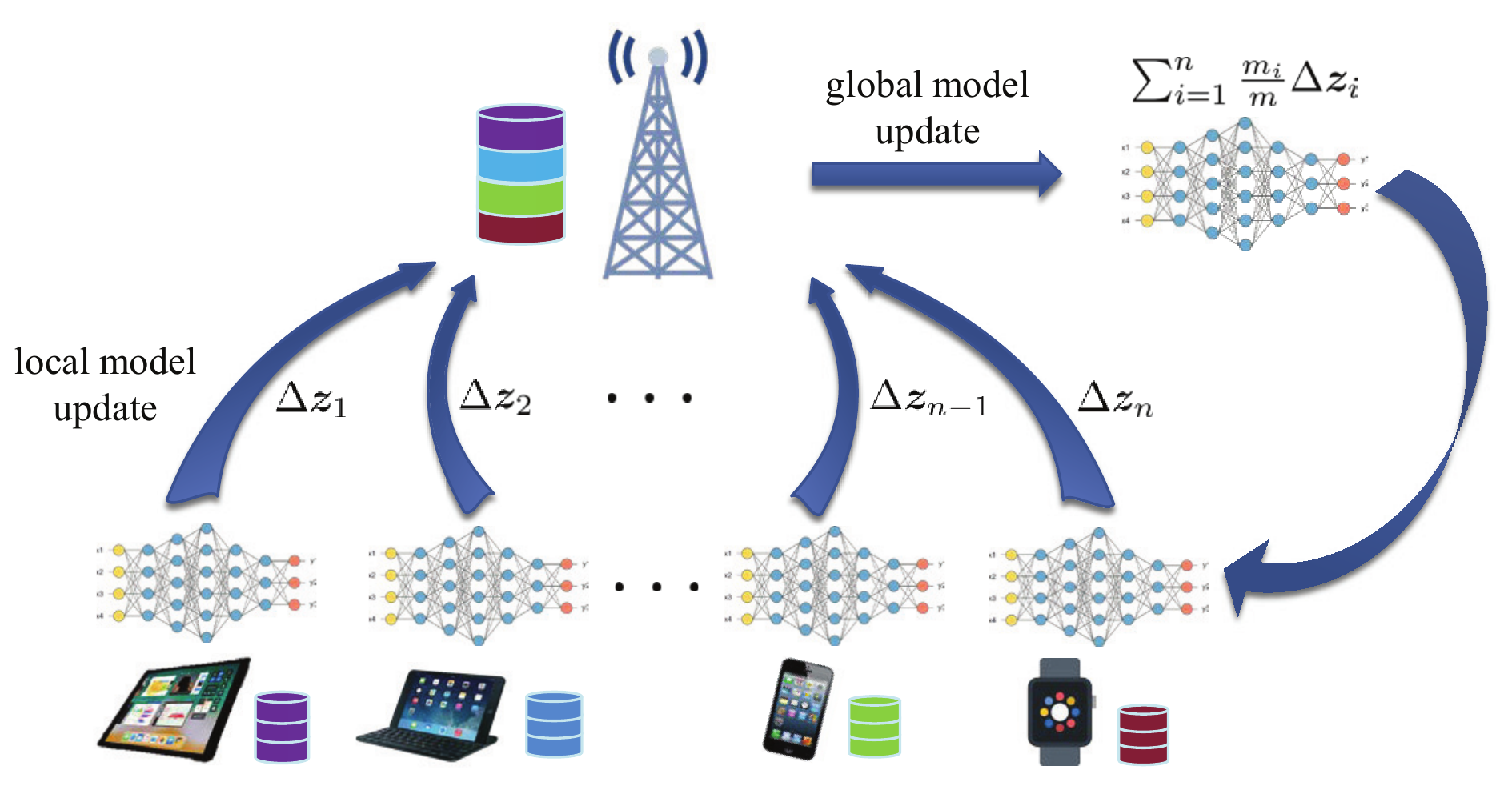}
\caption{Over-the-air computation for on-device distributed federated learning.}
\label{flair}
\end{figure}
Large-scale distributed machine learning is needed
for mobile AI applications in 6G, for which communication becomes the key bottleneck
for scaling up distributed training and distributed inference over the cloud,
network edge, and end-devices.

{\bf{Communication-Efficient Distributed Training}}:
The growing computation and storage power of devices provides opportunities for on-device distributed training by  processing data locally. However, communicating over the volatile wireless channel becomes the significant bottleneck for distributed training on mobile devices. To strengthen data privacy and security, federated learning  \cite{mcmahan2017communication} allows the training data to be kept at each device, thereby learning a shared global model from distributed mobile devices. However, the limited bandwidth becomes the main bottleneck for global model aggregation from locally updated models computed at each mobile device. The over-the-air computation can be exploited to enable low-latency global model aggregation via exploiting the superposition property
of a wireless multiple-access channel, as shown in Fig. {\ref{flair}}.

\begin{figure}[t]
\center
\includegraphics[width=0.85\columnwidth]{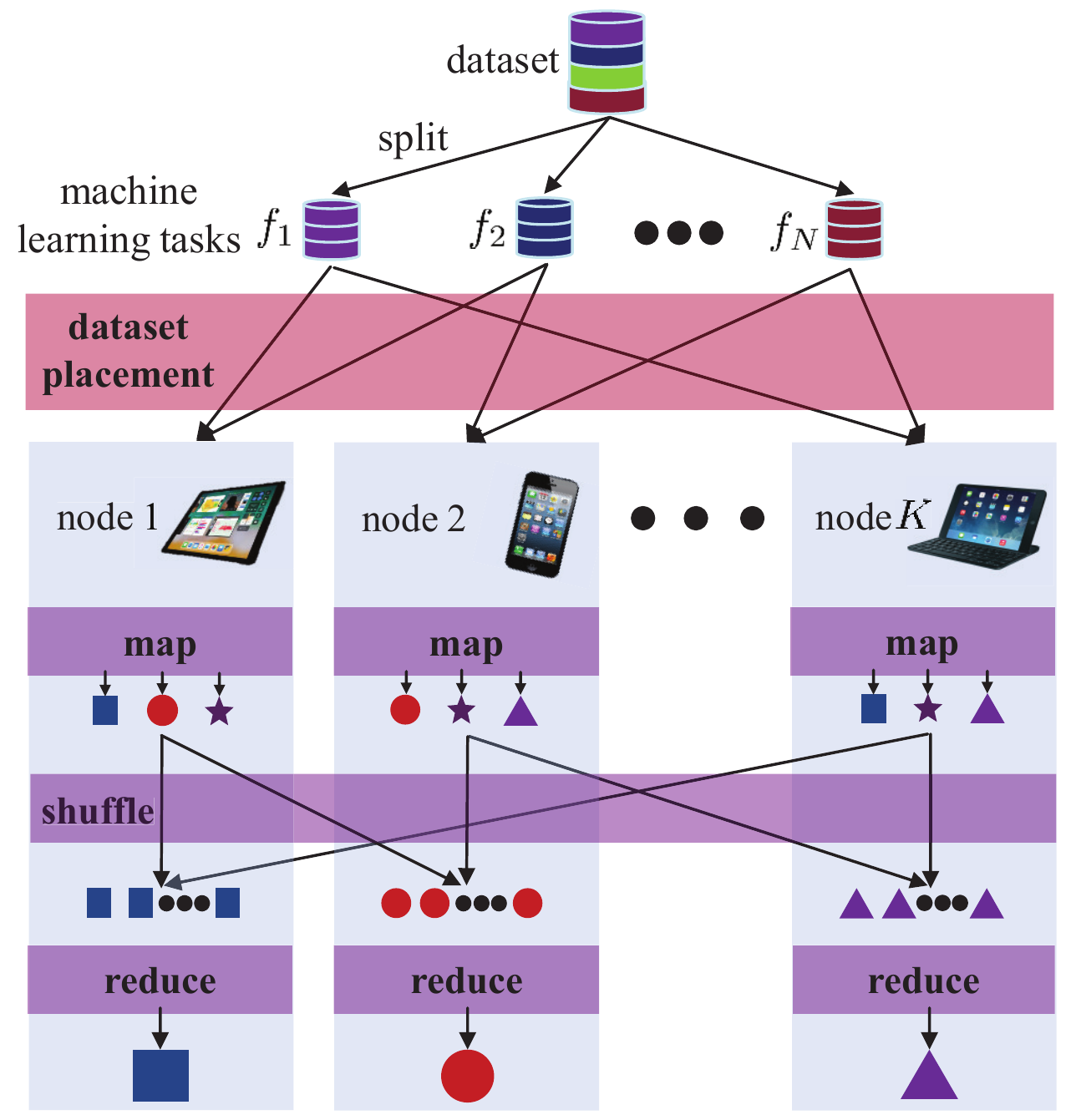}
\caption{On-device
distributed inference via wireless MapReduce.}
\label{mapreduce}
\end{figure}

{\bf{Communication-Efficient Distributed Inference}}:
In 6G, intelligent services will span  from cloud
data centers to end-devices and IoT devices, e.g., self-driving cars, drones, and auto-robots. As such, it is of prime importance to design ultra-low latency,  ultra-low power and low-cost
inference processes. To overcome  stringent computation,
bandwidth, storage, power and privacy constraints on individual devices, increasing research interests are moving towards
leveraging the dispersed computing resources across the cloud,
network edge and end-devices of 6G networks  through
the lens of mobile edge
computing \cite{Khaled_MEC17}. For example, for a deep neural network, the initial
features can be extracted on the end devices, which are then sent to the
edge and cloud computing devices for further processing. However, with the heterogeneity in the computing capabilities
and communication bandwidths among the computing devices, it becomes extremely
challenging to allocate the operations of the neural networks to the computing
devices so that the latency and energy are optimized. Fig. {\ref{mapreduce}} demonstrates the on-device distributed inference process, where each device locally computes the intermediate values based on the map function using the local data. The intermediate values are further shuffled across the devices assisted by a central radio access points. The inference process will be accomplished by collecting all the required intermediate values to construct the prediction results. A joint optimization of the uplink and downlink
communication strategy was thus developed in \cite{Yuanming_distcomp18icassp} for shuffling the locally computed intermediate values across mobile devices.

\section{Hardware-Aware Communications for 6G}
\label{hardwarecom}

\begin{figure}[t]
\center
\includegraphics[width=0.98\columnwidth]{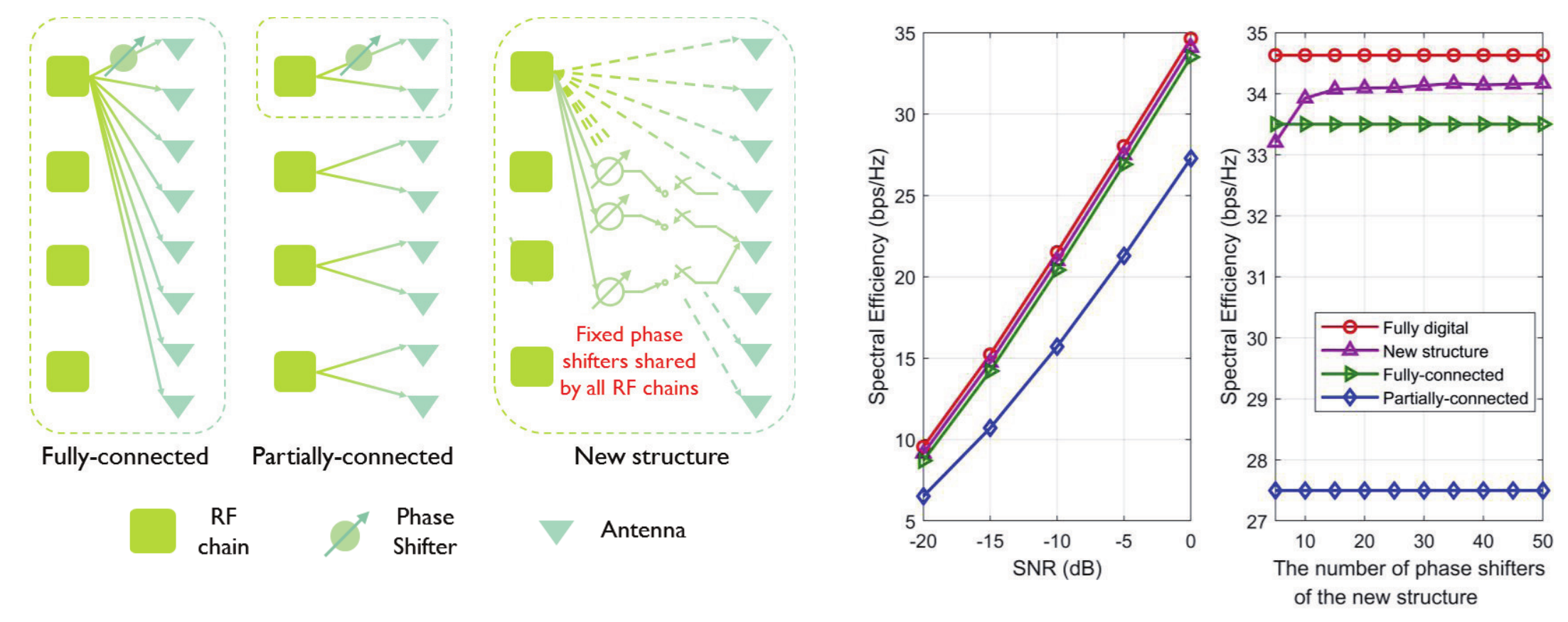}
\caption{A hardware-efficient hybrid beamforming structure with fixed phase
shifters. The base station and user are equipped with 144 and 16 antennas, respectively, and 4 RF chains. The fully- and partially-connected structures require 576 and 144 adaptive phase shifters, respectively, while the new structure only requires 30 fixed phase shifters in the first simulation. The second simulation shows that 15 phase shifters are already sufficient for the new structure.}
\label{hardware}
\end{figure}

As new radio access technologies emerge, and IoT devices become more pervasive, hardware constraints will play critical roles when designing 6G networks. On one hand, as radio communication is moving towards millimeter-wave bands, and possibly Terahertz bands, the high cost and power consumption of hardware components will significantly affect the transceiver architecture and algorithm design. On the other hand, IoT devices have limited storage, energy source, and on-device computing power. Such resource-constrained platforms call for a holistic design of communication, sensing, and inference. In this section, we present a new design paradigm for 6G, namely \emph{hardware-aware communications}, and discuss three promising new design principles. For performance-critical scenarios, the objective is to develop hardware-efficient transceivers that are also algorithm friendly, which calls for {\it{hardware-algorithm co-design}}. For IoT-like application scenarios, {\it{application-aware communications}} will be essential. Meanwhile, {\it{intelligent communications}} is needed to effectively adapt to heterogeneous hardware constraints.

\subsection{Hardware-Algorithm Co-design}
\label{haco}
The desire to communicate at ever higher data rates will never stop. To reach  Terabytes per second data rates, it is inevitable to operate at higher and higher frequency bands. The major obstacle is from the hardware perspective. Very large scale antenna arrays are needed to overcome the increased pathloss and other propagation phenomena, which will bring a large number of hardware components, including signal mixers, ADCs/DACs, power amplifiers, etc. The high cost and power consumption of these components at the mmWave and THz band make it difficult to adopt conventional transceiver structures, which in turn will affect the design of signal processing algorithms. To effectively design such complex systems, collaboration among the hardware and algorithm domains will be needed, i.e., hardware-algorithm co-design should be advocated. The target is to develop hardware-efficient transceiver structures that are also algorithm friendly: such structures should employ few of the costly hardware components, and they should be able to leverage existing signal processing algorithms.

\subsubsection*{Case Study}
Consider mmWave hybrid beamforming as an example, which is a cost-effective approach for providing effective beamforming gains. It requires a small number of RF chains, and thus can significantly reduce hardware cost and power consumption. However, a large number of phase shifters are still needed for existing hardware structure. Phase shifters at mmWave bands are still very expensive, and thus  their number needs to be reduced. A new hardware-efficient hybrid structure was recently proposed in \cite{Khaled_JSTSP18}, as shown in Fig. {\ref{hardware}}. It only requires a small number of phase shifters, each with a fixed phase. As such, hardware modification is only in the analog network, basic design principles for hybrid beamforming can still be applied. As shown in Fig. {\ref{hardware}}, this new structure can approach the performance of the fully digital beamforming, with much fewer phase shifters than other hybrid beamforming structures.

\subsection{Application-Aware Communications for IoT Devices}
Thanks to the recent development of IoT technologies, intelligent mobile applications will thrive, and many of them are powered by specialized low-cost, low-power devices. Such devices will handle basic sensing and simple on-device processing tasks, while relying on proximate edge servers or remote cloud data centers for computation-intensive processing. Thus, effective communications between devices and servers will be essential. Rather than serving as a bit pipe for traditional data services and focusing on maximizing data rates, wireless communications for IoT applications should directly serve specific applications. An integrated consideration of communication, sensing, and inference will be critical to overcome the hardware limitations, as illustrated below.

\subsubsection*{Joint sampling, communication, and inference}
IoT devices have serious challenges.  These include, 1) limited computing power to process the collected data; 2) their limited energy will  constrain their ability to collect data samples; 3) they do not have enough storage to store all the data; and 4) they cannot afford to always send data to the server. By jointly optimizing sampling, communication, and local processing, and accounting for the state of local processors, storage, and channel states, the overall performance can be improved. The integration with edge computing will play an important role, and joint edge-device processing techniques should be developed.


\subsection{Intelligent Communications for Heterogeneous Hardware Constraints}
Wireless networks are getting more and more heterogeneous, with various types of access points and mobile terminals, which differ significantly in hardware settings. Such heterogeneity has started from 4G LTE networks, and with the deployment of advanced techniques such as massive MIMO, the situation will further develop through 5G, and into 6G. This trend will complicate the communication protocol and algorithm design, which may subsequently degrade the communication efficiency. Recently, adopting machine learning techniques to develop communication systems has demonstrated its effectiveness, and such approaches have the potential of leading to general purpose intelligent communications that can adapt to heterogeneous hardware constraints. A particular approach is illustrated as follows.

\subsubsection*{Transfer learning for different hardware constraints}
One complication brought by hardware heterogeneity is the excessive effort to redesign the system for different hardware settings. For example, different transceiver architectures have been proposed for mmWave systems, including analog beamforming, hybrid beamforming, and 1-bit digital beamforming. The conventional approach relies on hand-crafted design for each of them, which is very inefficient. These different types of transceivers will face the same problems as those in mmWave channels, and thus an algorithm well designed for one may also shed light on the design for another. Transfer learning is a promising technique that can help to transfer the design of one architecture to others.


\section{Conclusions}
\label{conclud}
This article has presented an AI empowered architecture, as well as AI-centric communication techniques, for 6G networks.  New features of the 6G evolution were
identified, and enabling technologies were discussed.
While a partial picture was presented, we hope
our discussion will spur interests and further investigations on
the future evolution of cellular networks.

\bibliographystyle{ieeetr}

\begin{thebibliography}{10}

\bibitem{Fettweis_lowlatency18}
K.~{Chen}, T.~{Zhang}, R.~D. {Gitlin}, and G.~{Fettweis}, ``Ultra-low latency
  mobile networking,'' {\em IEEE Network}, pp.~1--7, 2018.

\bibitem{Jeff_JSAC5G}
J.~Andrews, S.~Buzzi, W.~Choi, S.~Hanly, A.~Lozano, A.~Soong, and J.~Zhang,
  ``What will 5{G} be?,'' {\em IEEE J. Sel. Areas Commun.}, vol.~32,
  pp.~1065--1082, Jun. 2014.

\bibitem{Berndt_6G}
K.~{David} and H.~{Berndt}, ``{6G} vision and requirements: Is there any
need
  for beyond {5G}?,'' {\em IEEE Veh. Technol. Mag.}, vol.~13, pp.~72--80,
Sep.
  2018.

\bibitem{strinati20196g}
E.~C. Strinati, S.~Barbarossa, J.~L. Gonzalez-Jimenez, D.~Kt{\'e}nas,
  N.~Cassiau, and C.~Dehos, ``{6G}: The next frontier,'' {\em arXiv preprint
  arXiv:1901.03239}, 2019.

\bibitem{Debbah_arXiv196G}
F.~Tariq, M.~Khandaker, K.-K. Wong, M.~Imran, M.~Bennis, and M.~Debbah, ``A
  speculative study on {6G},'' {\em arXiv:1902.06700}, 2019.

\bibitem{Rappaport_Globecom18}
Y.~{Xing} and T.~S. {Rappaport}, ``Propagation measurement system and approach
  at 140 {GH}z-moving to 6{G} and above 100 {GH}z,'' in {\em Proc. IEEE Global
  Communications Conf. (GLOBECOM)}, pp.~1--6, Dec. 2018.

\bibitem{Gesbert_IoT19}
O.~{Esrafilian}, R.~{Gangula}, and D.~{Gesbert}, ``Learning to communicate
in
  {UAV}-aided wireless networks: Map-based approaches,'' {\em IEEE Internet
of
  Things J.}, to appear, 2019.

\bibitem{Kato_WC19AI}
N.~{Kato}, Z.~{Md. Fadlullah}, F.~{Tang}, B.~{Mao}, S.~{Tani}, A.~{Okamura},
  and J.~{Liu}, ``Optimizing space-air-ground integrated networks by artificial
  intelligence,'' {\em IEEE Wireless Commun.}, pp.~1--8, 2019.

\bibitem{silver2017mastering}
D.~Silver, J.~Schrittwieser, K.~Simonyan, I.~Antonoglou, A.~Huang, A.~Guez,
  T.~Hubert, L.~Baker, M.~Lai, A.~Bolton, {\em et~al.}, ``Mastering the game
of
  go without human knowledge,'' {\em Nature}, vol.~550, no.~7676, p.~354,
2017.

\bibitem{Mitola_IPC19}
J.~Mitola and G.~Q. Maguire, ``Cognitive radio: making software radios more
  personal,'' {\em IEEE Pers. Commun.}, vol.~6, pp.~13--18, Aug. 1999.

\bibitem{Liang_DRLarXiv18}
N.~C. Luong, D.~T. Hoang, S.~Gong, D.~Niyato, P.~Wang, Y.~Liang, and D.~I.
Kim,
  ``Applications of deep reinforcement learning in communications and
  networking: {A} survey,'' {\em CoRR}, vol.~abs/1810.07862, 2018.

\bibitem{mcmahan2017communication}
B.~McMahan, E.~Moore, D.~Ramage, S.~Hampson, and B.~A. y~Arcas,
  ``Communication-efficient learning of deep networks from decentralized
  data,'' in {\em Proc. Int. Conf. Artificial Intell. Stat. (AISTATS)},
  vol.~54, pp.~1273--1282, 2017.

\bibitem{Yuanming_distcomp18icassp}
K.~Yang, Y.~Shi, and Z.~Ding, ``Low-rank optimization for data shuffling
in
  wireless distributed computing,'' in {\em Proc. IEEE Int. Conf. Acoustics
  Speech Signal Process. (ICASSP)}, Calgary, Alberta, Canada, 2018.

\bibitem{Khaled_MEC17}
Y.~Mao, C.~You, J.~Zhang, K.~Huang, and K.~B. Letaief, ``A survey on mobile
  edge computing: The communication perspective,'' {\em IEEE Commun. Surveys
  Tuts.}, vol.~19, pp.~2322--2358, Fourth quarter, 2017.

\bibitem{Khaled_JSTSP18}
X.~Yu, J.~Zhang, and K.~B. Letaief, ``A hardware-efficient analog network
  structure for hybrid precoding in millimeter wave systems,'' {\em IEEE
J.
  Sel. Topics Signal Process.}, vol.~12, pp.~282--297, May 2018.

\end{thebibliography}

 {\bf{Khaled B. Letaief}} [S'85-M'86-SM'97-F'03] (eekhaled@ust.hk) received his Ph.D. degree
from Purdue University. From 1990 to 1993, he was a faculty member at the
University of Melbourne, Australia. He has been with HKUST since 1993 where
he was Dean of Engineering, and is now the New Bright Professor of Engineering.
From 2015 to 2018, he joined HBKU in Qatar as Provost. He is an ISI Highly
Cited Researcher and a recipient of many distinguished awards. He has served
in many IEEE leadership positions including ComSoc President (at present),
Vice-President for Technical Activities, and Vice-President for Conferences.
\\[-3mm]

 {\bf{Wei Chen}} [S'05-M'07-SM'13] (wchen@tsinghua.edu.cn) received his B.S. and Ph.D. degrees  from Tsinghua University. He was a visiting Ph.D. student at HKUST from 2005 to 2007. He is currently a tenured Professor at the Department  of Electronic
Engineering, Tsinghua University. \\[-3mm]

{\bf{Yuanming Shi}} [S'13-M'15] (shiym@shanghaitech.edu.cn) received his B.S.
 degree  from Tsinghua University and the Ph.D. degree
from  The Hong Kong University of Science and Technology.
He is currently a tenured Associate Professor at the School of Information
Science and Technology, ShanghaiTech University. \\[-3mm]

{\bf{Jun Zhang}} [S'06-M'10-SM'15] (jun-eie.zhang@polyu.edu.hk) received his Ph.D. degree from the University of Texas at Austin. He is currently an Assistant Professor at The Hong Kong Polytechnic University. \\[-3mm]

{\bf{Ying-Jun Angela Zhang}} [S'00-M'05-SM'10] (yjzhang@ie.cuhk.edu.hk) received her Ph.D. degree from The Hong Kong University of Science and Technology. She is now an Associate Professor at the Department of Information Engineering, The Chinese University of Hong Kong.

\end{document}